\documentclass[12pt,preprint]{aastex}

\def\gtsim{\ {\raise-0.5ex\hbox{$\buildrel>\over\sim$}}\ }
\def\ltsim{\ {\raise-0.5ex\hbox{$\buildrel<\over\sim$}}\ }

\begin{document}

\title{ Millimagnitude Photometry for Transiting Extrasolar Planetary
Candidates:~II. Transits of OGLE-TR-113-b in the Optical and Near-IR
\footnote{Based on observations collected with the 
Very Large Telescope at Paranal Observatory 
(JMF and DM visiting observers), and at the
ESO New Technology Telescope at La Silla Observatory (SR, FP and DM visiting
observers) for the ESO Programmes 075.C-0427, 075.B-0414, and 076.C-0122.
}}

\author{
Rodrigo F.~D\'\i az\altaffilmark{1},
Sebasti\'an Ram\'{\i}rez\altaffilmark{2},
Jos\'e Miguel Fern\'andez\altaffilmark{2,8},
Jos\'e Gallardo\altaffilmark{7},
Wolfgang Gieren\altaffilmark{3},
Valentin D. Ivanov\altaffilmark{6},
Pablo Mauas\altaffilmark{1},
Dante Minniti\altaffilmark{2},
Grzegorz Pietrzynski\altaffilmark{3,5},
Felipe P\'erez\altaffilmark{2},
Mar\'ia Teresa Ru\'iz\altaffilmark{4},
Andrzej Udalski\altaffilmark{5} and
Manuela Zoccali\altaffilmark{2}
}

\altaffiltext{1}{Instituto de Astronom\'{\i}a y F\'{\i}sica del
Espacio, CONICET-Universidad de Buenos Aires, C.C. 67 Suc. 28 (1428),
Buenos Aires, Argentina \\ E-mail: rodrigo@iafe.uba.ar,
pablo@iafe.uba.ar}

\altaffiltext{2}{Department of Astronomy, Pontificia Universidad Cat\'olica, 
Casilla 306, Santiago 22, Chile\\
E-mail:  sramirez@astro.puc.cl, dante@astro.puc.cl,
fperez@astro.puc.cl, mzoccali@astro.puc.cl}

\altaffiltext{3}{Department of Physics, Universidad de Concepci\'on,
Casilla 160-C, Concepci\'on, Chile \\ E-mail:
pietrzyn@hubble.cfm.udec.cl, wgieren@astro-udec.cl}

\altaffiltext{4}{Department of Astronomy, Universidad de Chile, Santiago, Chile
\\
E-mail: mtruiz@das.uchile.cl}

\altaffiltext{5}{Warsaw University Observatory, Al. Ujazdowskie 4, 00-478 Waszawa, Poland
\\
E-mail: udalski@astrouw.edu.pl}

\altaffiltext{6}{European Southern Observatory, Vitacura, Santiago, Chile
\\
E-mail: vivanov@eso.org}

\altaffiltext{7}{{CNRS UMR 5574, CRAL, \'Ecole normale sup\'erieure, 69364, Lyon Cedex 07, France\\
E-mail: jose.gallardo@ens-lyon.fr}}

\altaffiltext{8}{Harvard CFA, USA
\\
E-mail: jfernand@cfa.harvard.edu}

\begin{abstract}

We present precise V and $K_s$-band transit photometry for the
planetary host star OGLE-TR-113. Using the $K_s$-band photometry, we
confirm the dwarf nature of OGLE-TR-113, and obtain new estimates for
its effective temperature, distance and reddening. We employ the
V-band photometry to obtain planetary and orbit parameters from the
transit fit, $\mathrm{a}=(0.0232\pm0.0038)$ AU, orbital period
$\mathrm{P}=(1.4324752\pm0.0000015)$~days, $i=86.7-90$,
$\mathrm{R_p}=(1.09\pm0.09)~\mathrm{R_J}$. These values are in
excellent agreement with previous works. Assuming a mass
$\mathrm{M_p}=(1.32\pm0.19)~\mathrm{M_J}$ for the planet we obtain its
mean density $\rho=(1.26\pm0.50)~\mathrm{g~cm^{-3}}$, also in
agreement with previous works. The transit observed in the $K_s$-band
has a larger scatter and we find its amplitude to be consistent with
that in the V-band. In this way, we find an independent confirmation
of the planetary nature of OGLE-TR-113b.

\end{abstract}


\section{Introduction}
The discovery of hot Jupiters that transit in front of their parent
stars has advanced our knowledge of extrasolar planets adding a
fundamental datum: the planetary radius. Besides, combined with radial
velocity measurements, it allows a precise measurement of the
companion mass, and therefore, its mean density. This observational
data are fundamental for the development of models. New samples of
transiting hot Jupiters should become available soon (\citealp[see,
for example,][]{borde2003} (COROT); \citealp[][]{alonso2004} and
\citealp{o'donovan2006} (TrES); \citealp[][]{bakos2004} (HAT);
\citealp{fischer2005} (N2K); \citealp[][]{pollacco2006} (WASP);
\citealp[][]{mccullough2005} (XO); \citealp{aigrain2006} and
\citealp{irwin2006} (Monitor); \citealp{holman2006} (TLC);
\citealp{sahu2006} (SWEEPS)) but up to now the Optical Gravitational
Lensing Experiment (OGLE) has provided the largest number of
transiting candidates
\citep{udalski2002a,udalski2002c,udalski2002b,udalski2003}. In
particular, \citet{udalski2002c} discovered transits in the
$\mathrm{I}=14.42$ magnitude star OGLE-TR-113, located in the Carina
region of the Milky Way disk, at
$\mathrm{RA(2000)=10^h52^m24^{\prime\prime}.40,~DEC(2000)=-61^\circ
26^{\prime} 48^{\prime\prime}.5}$. They monitored 10 individual
transits, measuring an amplitude $\Delta\mathrm{I}=0.030$ mag, and a
period $\mathrm{P}=1.43250$ days.

The planetary nature of the transiting candidate OGLE-TR-113b was
confirmed by \citet{bouchy2004} and \citet{konacki2004} using high
precision radial velocities measurements. The results are shown in
Table~\ref{resultados} and confirm that OGLE-TR-113b can be classified
as a Very Hot Jupiter. Recently, \citet{gillon2006} and
\citet{snellencovino2006} presented R and K-band transit photometry,
respectively. They obtain parameters in excellent agreement with
previous works. Additionally, \citet{snellencovino2006} report a
tentative detection of emission from the planet.

There are many difficulties to confirm real planets among the low
amplitude transit objects discovered by photometric transit
searches. Most of the candidates turn out to be grazing binaries, low
mass stars, triples or blends \citep[see, for example,][]{torres2005,
pont2005, mandushev2005, o'donovan2006a}. Furthermore, the red noise
sometimes causes false detections \citep{pont2006}. Therefore, real
planets must be confirmed with accurate radial velocities, which in
the case of the OGLE candidates is difficult because their magnitudes
range from V\,=\,15 to 18 and they are located in very crowded fields.

Another test to check the real planetary nature of low
amplitude transiting objects is to observe their transits in different
wavelengths.  To first order, ignoring stellar limb darkening, the
transit amplitudes should be achromatic, i.e. they should be similar
in the V, I, and $K_s$-bands. Large amplitude differences would be
indicative of blends.

A triple system where a low mass star transits in front of a solar
type star that is blended with a foreground red star would give larger
amplitudes in the $K_s$-band. On the other hand, if the blend is with
an early type star in the foreground, the amplitudes would be larger
in the V-band. Unfortunately, blends with foreground stars of similar
color as the eclipsed star would yield similar transit depths in all
bands.  In this sense, while multicolor transit photometry can help to
discriminate most impostors, it cannot be regarded as final proof of
the true nature of the planetary companion \citep[see][]{moutou2005}.

To second order, the multiple wavelength observations allow to test
limb darkening coefficients. This is so because the shapes of the
transit light curves vary with wavelength due to stellar limb
darkening and to a possible planetary atmosphere \citep{burrows2000}.

We have undertaken an observational project at the ESO telescopes at
La Silla and Paranal Observatories, to measure transit depths in the
$K_s$ and V-bands, to compare with the OGLE I-band observations. In the
first paper we presented optical photometry of the planetary transit
candidate OGLE-TR-109b (\citeauthor{jm2006}
2006, hereafter Paper I)\defcitealias{jm2006}{Paper I}. Other papers
in the series discuss OGLE-TR-111 \citep{minniti2007} and OGLE-TR-82
\citep{hoyer2007}.

In this work we present new photometry covering a transit of
OGLE-TR-113b in the V-band with the ESO VLT, and another transit in
the $K_s$-band with the ESO NTT, improving the measurements of the
planetary parameters. The present observations allow to check
independently and to refine the ephemeris and amplitudes, and to
determine variations on the radius measurement using different
photometric filters.

There is a fundamental difference in our observations with respect to
OGLE. The OGLE candidates have light curves made of several
transits. While there are relatively few OGLE I-band points per
transit, the final phased light curves are very clean, and
representative of an average transit. Our data show single transits
but very well sampled. As such, they could in principle be sensitive
to large planetary satellites, for example \citep{charbonneau2004}.

This paper is organized as follows. Sections 2 and 3 present the infrared
and optical observations and photometry, respectively. Section 4 discusses
the parameters of the target derived from the present observations. Finally, 
the conclusions are outlined in Section 5.

\section{IR Observations and Photometry}
\subsection {IR Observations and Data Reduction}

\begin{figure}
\plotone{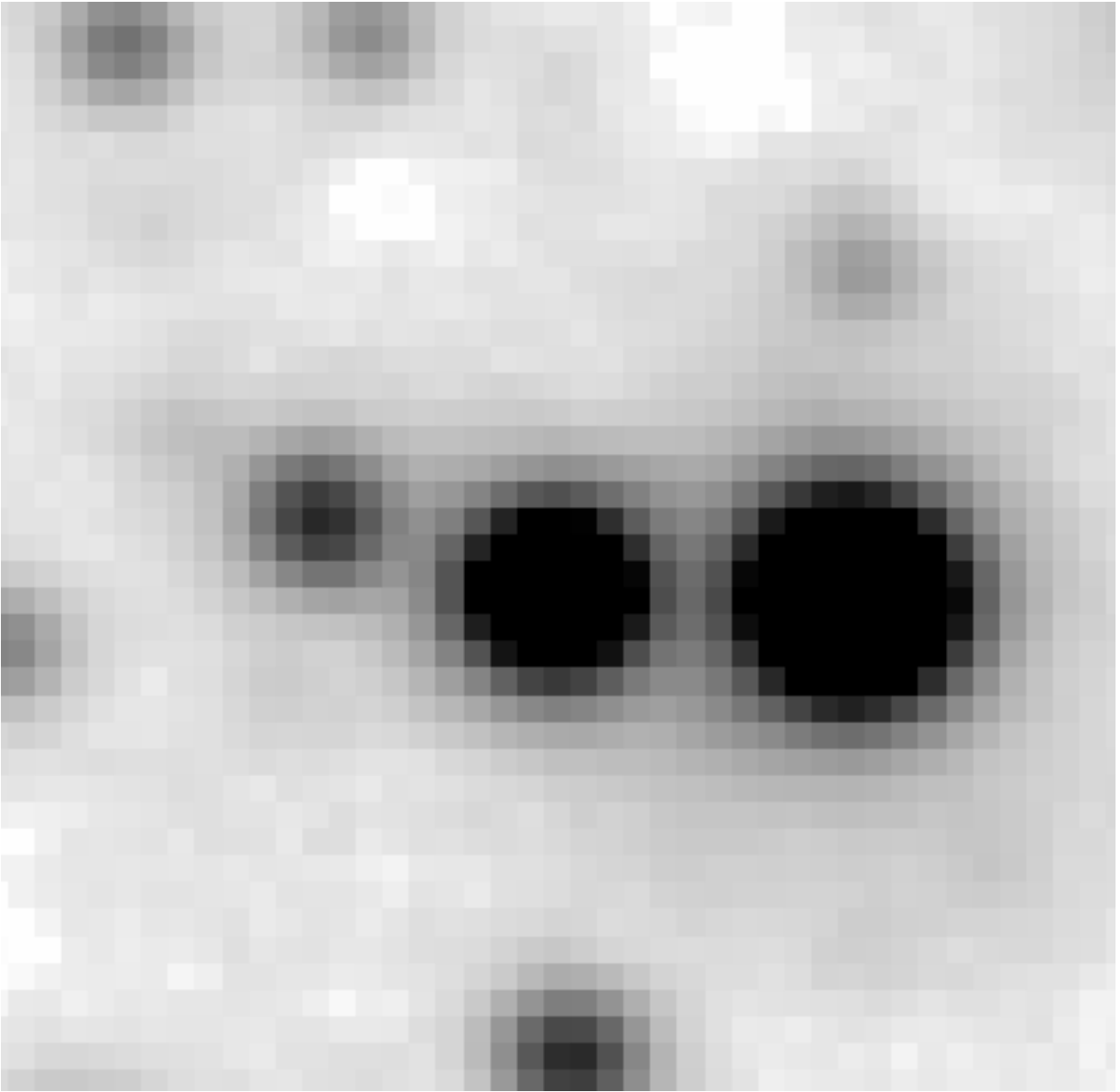}
\caption{Portion of a SOFI $K_s$-band image including OGLE-TR-113
($K_s=13.00$), which is the bright star at the center. This image
covers $12\times 12$ arcsec, and the faintest stars seen have $K_s\sim
18$.}
\label{K_image}
\end{figure}

OGLE-TR-113 was observed during the nights of May 4 and 5, 2005, using
the SOFI IR camera and spectrograph at the ESO NTT. SOFI is equipped with a Hawaii
HgCdTe detector of 1024$\times$1024 pixels, characterized by a 5.4
e/ADU gain, a readout noise of 2.1 ADU and a dark current of less than 0.1
e/sec. We used it in the Large Field camera mode, giving a 4.9 $\times$
4.9 arcmin$^2$ field.  All measurements were made through the
$K_s$-band filter ($\lambda_0 = 2.162$ $\mu$m and $\Delta\lambda =
0.275$ $\mu$m). In Figure~\ref{K_image} we show a $12\times 12$
arcsec$^2$ portion of an image acquired with SOFI.

We monitored several OGLE candidates (OGLE-TR-108, OGLE-TR-109,
OGLE-TR-113, OGLE-TR-170, and OGLE-TR-171), being in some cases
limited by the weather (high wind and clouds). The telescope was
slightly defocussed to avoid saturation of the detector for the
brightest stars. This does not affect the differential aperture
photometry nor our final results.

The reductions were made using IRAF tasks
\footnote{IRAF is distributed by the National Optical Observatories,
operated by Universities for Research in Astronomy, Inc.}.  First of
all, the crosstalk correction was applied, taking into account the
detectors sensitivity difference between the upper and lower
half. Then the sky substraction was applied.  The whole dataset was
acquired using "dither-5" around two offset positions that included
the target. These contiguous sky images were used to generate local
skies close in time for each of the offset images. Then the
appropriately scaled skies were substracted from the images. Finally,
we applied flat-field corrections to all images and aligned them. For
the flat-fields, we used the correction images provided by the NTT
SciOps
team\footnote{www.ls.eso.org/lasilla/sciops/ntt/sofi/reduction/flat\_fielding.html},
and the alignment was done with \emph{lintran} and \emph{imshift}.

\begin{figure}
\plotone{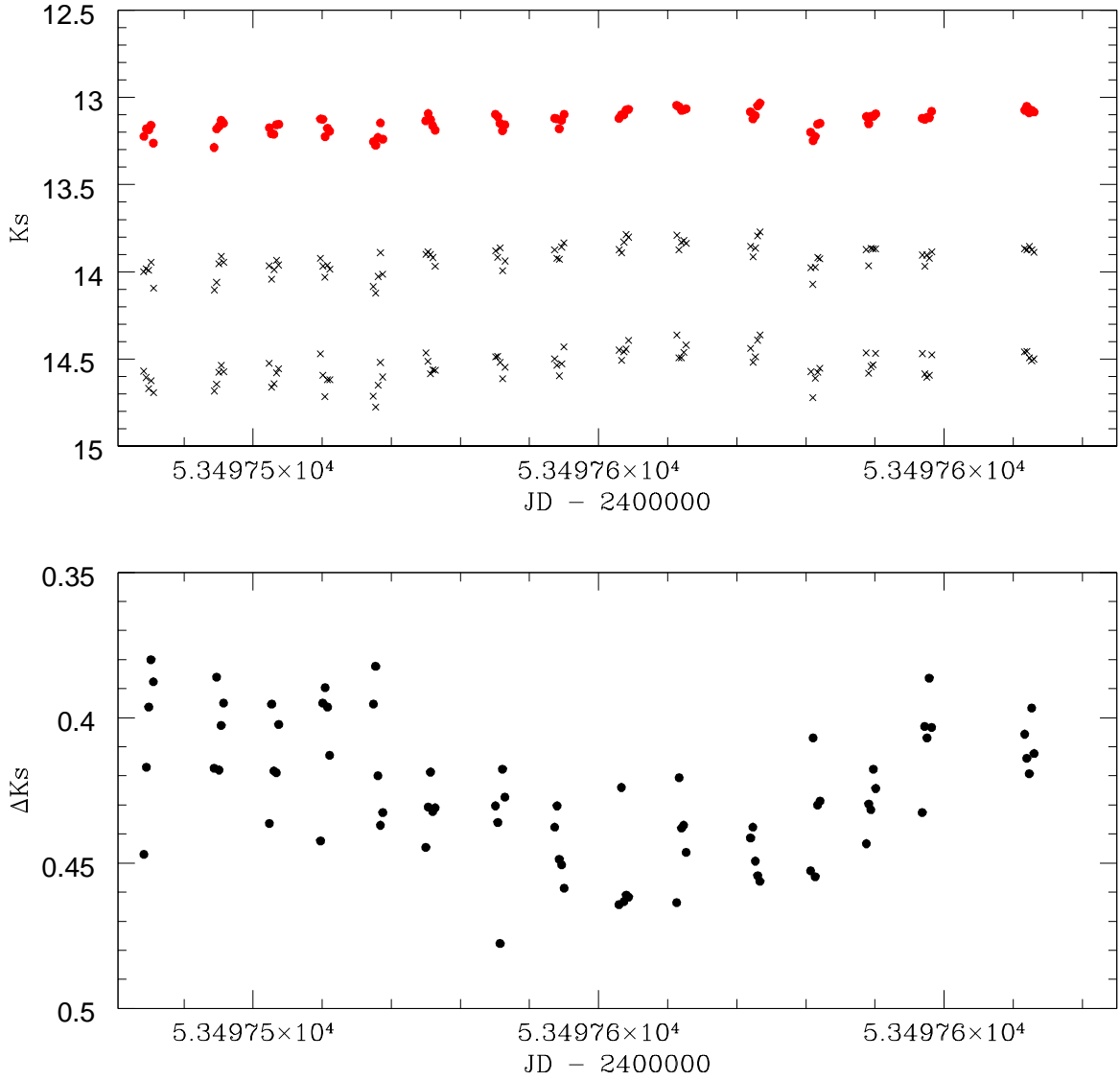}
\caption{Top: Light curves of OGLE-TR-113 (circles) and two
comparison stars (crosses) in the $K_s$-band. Bottom: Transit in the
$K_s$-band.
\label{kphoto}
}
\end{figure}

The calibration of the IR photometry was made using 2MASS. There were
a dozen stars in common located in the OGLE-TR-113 field (not
including this star because it is not listed as a 2MASS source).
Discarding outliers, we selected the 10 most isolated stars with
$12<K_s<14.5$, obtaining $\mathrm{RMS}_K=0.03$. The zero point of the
$K_s$-band photometry should be good to 0.1 mag, which is accurate
enough for our purposes. We also checked this calibration against the
DENIS sources in the field, finding excellent agreement. Finally, we
find $K_s=13.0\pm 0.1$ and $\mathrm{V}-K_s=3.5\pm 0.1$ for this
star. The resulting light curve is shown in Figure~\ref{kphoto}.

\subsection{IR photometry}
\subsubsection*{Stellar parameters\label{ir_phot}} 
Based on the high dispersion spectroscopy, \citet{konacki2004}
classify OGLE-TR-113 as a K-type main sequence star with
T$_{\mathrm{eff}} = 4800\pm 150$ K, gravity $\log~g = 4.5$, and
metallicity $\mathrm{[Fe/H]}=+0.0$ dex. They estimate the star is
located at a distance of about 600 pc. They adopt for this star a mass
of M$_*= 0.79\pm 0.06~ $M$_{\odot}$, and a radius R$_*=0.78\pm 0.06~
$R$_{\odot}$, very similar to the more recent values presented by
\citet{gillon2006}, who used spectroscopic measurements to obtain the
primary mass (M$_*= 0.78\pm 0.02$ M$_\odot$), and R-band high
precision transit photometry to obtain the stellar radius
(R$_*=0.77\pm 0.02$ R$_\odot$). Recently, \citet{santos2006} derived
similar stellar parameters for OGLE-TR-113: temperature
T$_{\mathrm{eff}} = 4804\pm 106~ $K, gravity $\log ~g = 4.52 \pm
0.26$, metallicity $\mathrm{[Fe/H]}=+0.15\pm0.10$ dex, based on high
dispersion spectroscopy.  They also derived the distance
$\mathrm{d}=553$ pc, and an absorption $A_V=0.42$ based on the OGLE
photometry. The values from these two independent studies agree within
the uncertainties.

With the present optical and IR photometry we can confirm
independently some of these parameters: the spectral type, luminosity,
mass, radius and distance.

\begin{figure}
\plotone{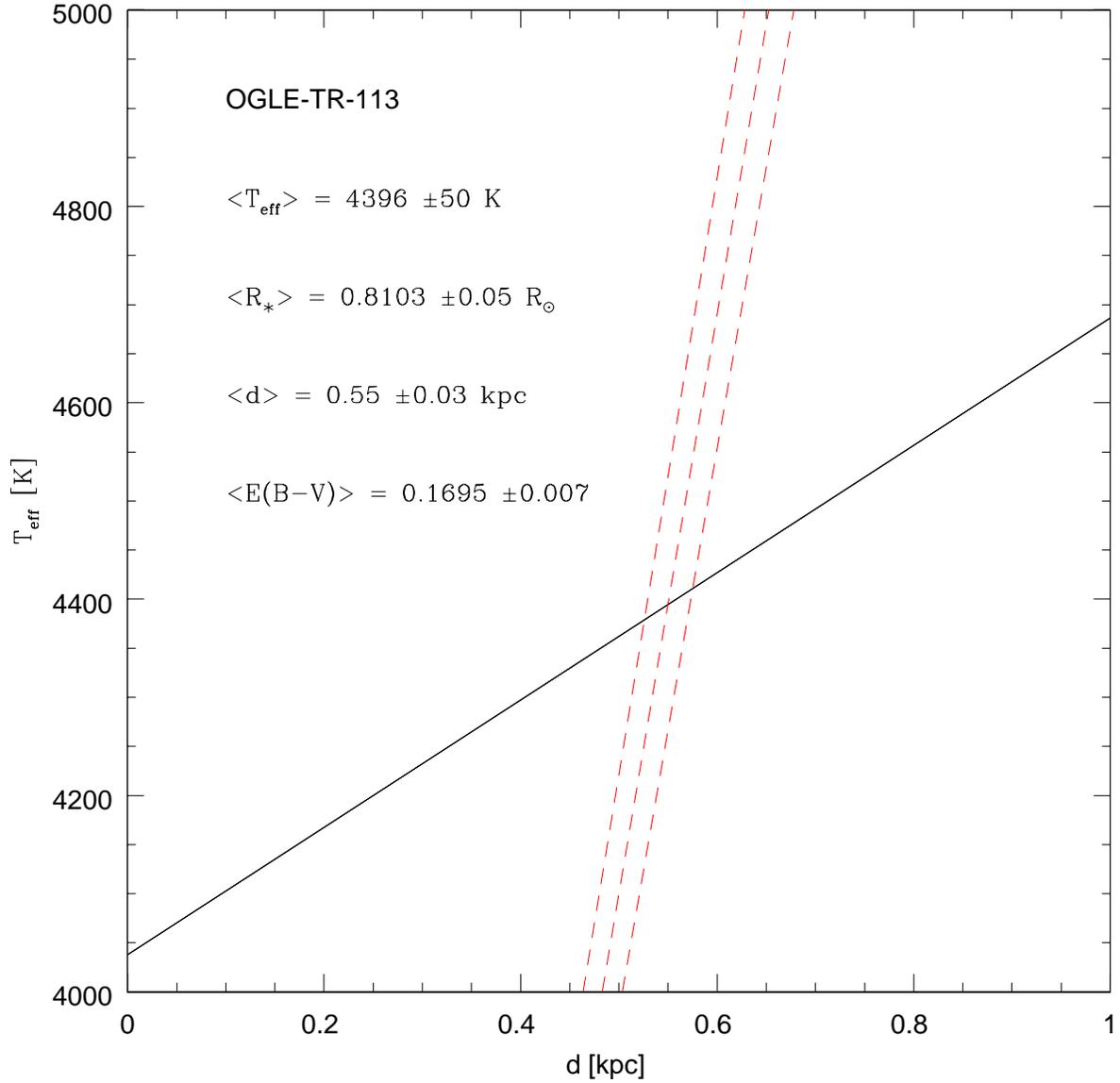}
\caption{ Effective temperature and distance for OGLE-TR-113b.  The
dashed (red) lines show the radius-temperature constrain for
main-sequence stars (errors in the optical and infrared photometry are
considered). Black lines give the temperature inferred from the
surface brightness relations (see \citealt[Section 4]{gallardo2005}
for further explication).
\label{parametros}
}
\end{figure}

In a previous work, we have used optical and infrared photometry to
characterize OGLE extrasolar planetary companions using surface
brightness analysis \citep{gallardo2005}. Unfortunately, at the time
we did not have $K_s$-band photometry of OGLE-TR-113. Following the
analysis described in that work, we obtained from the IR photometry
surface brightness the following stellar parameters: temperature
T$_{\mathrm{eff}}=4396\pm50$ K, radius R$_* =0.81\pm0.05$ R$_\odot$,
distance $\mathrm{d}=550\pm30$ pc, and reddening
$\mathrm{E_{(B-V)}}=0.17\pm0.01$ mag (see Figure~\ref{parametros}). In
spite of the lower temperature, the distance and reddening are similar
to those measured by \citet{santos2006}, and the radius is similar to
that of \citet{gillon2006}. The determined distance of 550 pc is
consistent with the $K_s$-band photometry, making this the nearest
known OGLE planet.  This is also the smallest OGLE star with a
confirmed planet.

\begin{figure}
\plotone{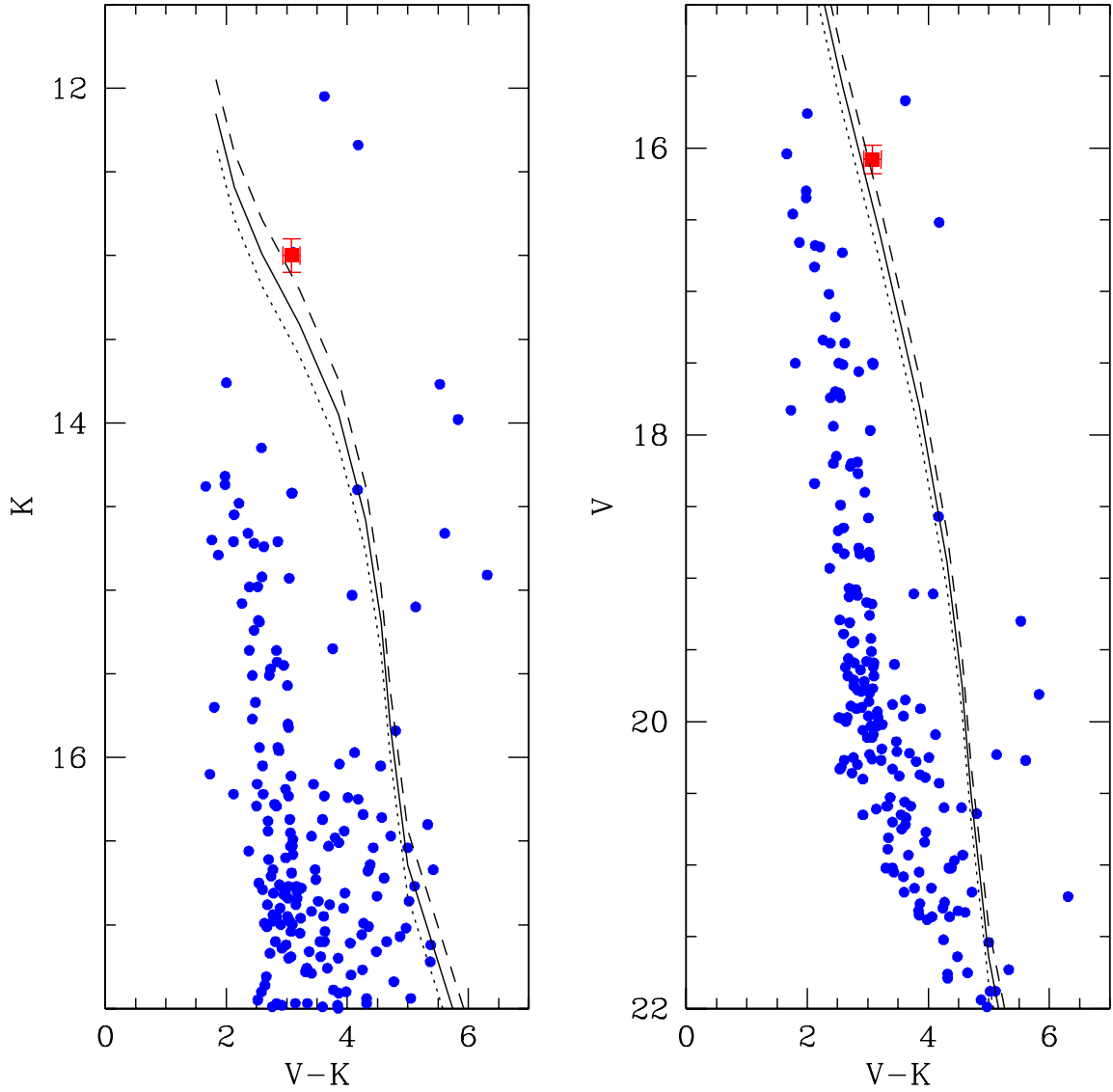}
\caption{ Optical-infrared color-magnitude diagrams of a $1'\times 1'$
field centered in OGLE-TR-113. The (red) filled square marks the
location of OGLE-TR-113. Isochrones for solar age and metallicity, for
different distances are also shown: $\mathrm{d}=550$ pc (solid line),
$\mathrm{d}=600$ pc (dotted line) and $\mathrm{d}=500$ pc (dashed
line).
\label{cmd}
}
\end{figure}

Figure \ref {cmd} shows the optical and IR CMDs for the stars
in a $1'\times 1'$ field centered on OGLE-TR-113. The filled red
square represents the extrasolar host star OGLE-TR-113. Additionally,
we show isochrones for solar age and metallicity for three different
distances, calculated using the value for colour excess and extinction
determined above \citetext{J. Gallardo, private communication}. The
disk main sequence is very well defined, and the position of the
target star is consistent with a dwarf star, despite being apparently
located away from the main sequence, indicating that either this is a
nearby K-dwarf, or a subgiant located farther than the majority of the
stars in this field. We prefer the first option based on the
independent evidence from the spectroscopy.  For example, a K0V star
has M$_{K_s}=3.9$, and a K5V star has M$_{K_s}=4.7$
\citep{allen}. These give $K_s=12.8$ and 13.6 for a distance of 600
pc, neglecting interstellar dust extinction.

\section{Optical Observations and Photometry}
\subsection{Optical Observations and data reduction}
The observations and photometry are described in detail in
\citetalias{jm2006}. Photometric observations were taken with VIMOS at the
Unit Telescope 4 (UT4) of the European Southern Observatory Very Large
Telescope (ESO VLT) at Paranal Observatory during the nights of April
9 to 12, 2005.  The VIMOS field of view consists of four CCDs, each
covering 7$\times$8 arcmin, with a separation gap of 2 arcmin, and a
pixel scale of 0.205 arcsec/pixel. Since the scope of the observations
is to detect new transit candidates, we monitored four different
fields. A number of OGLE transit candidates located in the observed
fields were monitored simultaneously. OGLE-TR-113 happened to have a
transit towards the end of the second night of our run.

Clearly, there exists a compromise between the temporal resolution of
the observations and the number of fields monitored. The observations
presented here were carried out alternating between two different
fields, observing three 15 sec exposures on each one. For this program
we managed to reduce the observation overheads for telescope presets,
instrument setups, and the telescope active optics configuration to an
absolute minimum. This ensured adequate sampling of the transit
providing at the same time $>10000$ stars with $15<V<19$ for which
lightcurves can be obtained. If individual candidates are to be
observed a much better temporal resolution can be achieved using other
telescopes or instruments (see \citet{gillon2006} for an example of
high-temporal resolution observations with NTT/SUSI and \citet{666_1}
for observations of OGLE-TR-10 and OGLE-TR-56 acquired with
VLT/FORS.) We obtained 150 points in the field of OGLE-TR-113 during
the second night. The observations lasted for about 9.5 hours, until
the field went below 3 airmasses.

We used the Bessell V filter of VIMOS, with $\lambda_0=5460$ \AA,
$\mathrm{FWHM}=890$ \AA. The V-band was chosen in order to
complement the OGLE light curves which are made with the I-band
filter.  In addition, the V-band is more sensitive to the effects of
limb darkening during the transit, and is adequate for the modeling of
the transit parameters.

\begin{figure}
\plotone{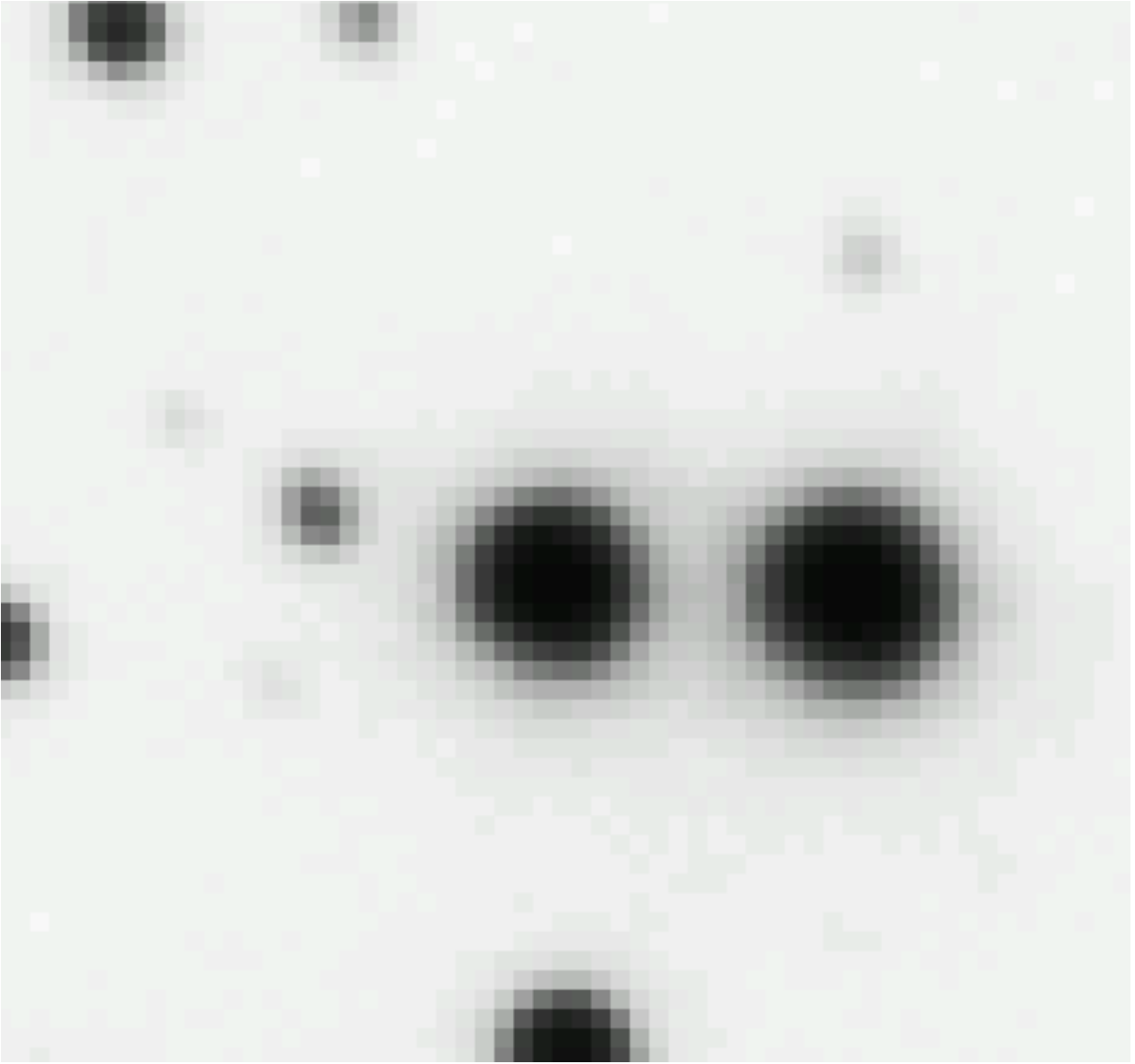}
\caption{Portion of a 0.6 arcsec seeing VIMOS image including
OGLE-TR-113 ($\mathrm{V}=16.08$), which is the bright star at the
center. This image covers $12\times 12$ arcsec as in
Figure~\ref{K_image}, and the faintest stars seen have $\mathrm{V}\sim24$.  }
\label{V_image}
\end{figure}

Figure~\ref{V_image} shows a $12\times 12$ arcsec$^2$ portion of a 0.6
arcsec seeing image, illustrating that the candidate star is
unblended. There are a $\mathrm{V}=16.0$ mag star 2 arcsec East and a
$\mathrm{V}=17.5$ star 2 arcsec West of our target.  These do not
affect our photometry.

\subsection{V-band transit photometry}
In order to reduce the analysis time of the vast dataset acquired with
VIMOS, the images of OGLE-TR-113 analyzed here are 400$\times$400 pix,
or 80 arcsec on a side. Each of these small images contains about 500
stars with $15<\mathrm{V}<24$ that can be used by the software ISIS for the
differential photometry, and in the light curve analysis. The best
seeing images ($\mathrm{FWHM}=0.6$ arcsec) taken near the zenith were selected,
and a master image was made in order to serve as reference for the
difference image analysis \citep[see][]{alard98,alard2000}.

The seeing, position of the stars, peak counts, and sky counts were
monitored. The individual light curves were checked against these
parameters in search for systematic effects.  Extensive tests with the
difference image photometry were performed, varying different
photometric parameters and choosing different sets of reference stars.
The software gives relative fluxes that are dependent on the reference
stars used, so that the final amplitude was measured using aperture
photometry in the individual images. To do this we selected three images
centered on the transit and three images before the transit acquired
with similar seeing but different airmass. Following this procedure we
believe that the measurement of the transit amplitude is reliable.

\begin{figure}
\plotone{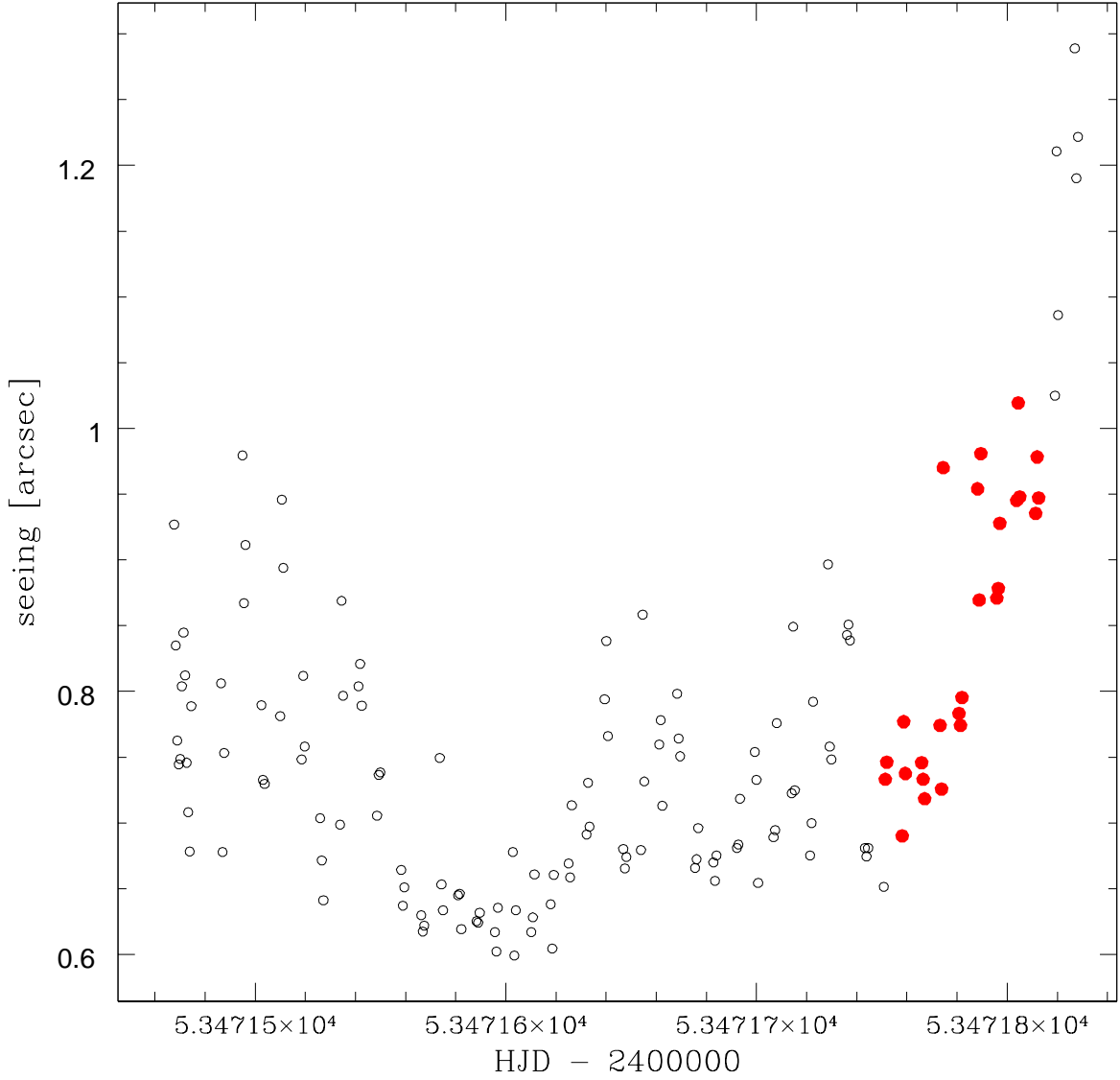}
\caption{
Seeing during the second observation night with VIMOS. The seeing
remained below 1 arcsecond for most of the night. The filled (red) points
correspond to the transit.
\label{fwhm}}
\end{figure}

The photometry of OGLE-TR-113, with mean $\mathrm{V}=16.08$ gives
$\mathrm{RMS_V}=0.0015$ to $0.0024$ mag throughout the second night of
the run. This quantity is found to correlate mostly with the sizes of
the point sources, given by a combination of seeing and
airmass. Therefore, the photometric scatter increased for the end of
the night, when the transit occurred (see Figure~\ref{fwhm}).

\begin{figure}
\plotone{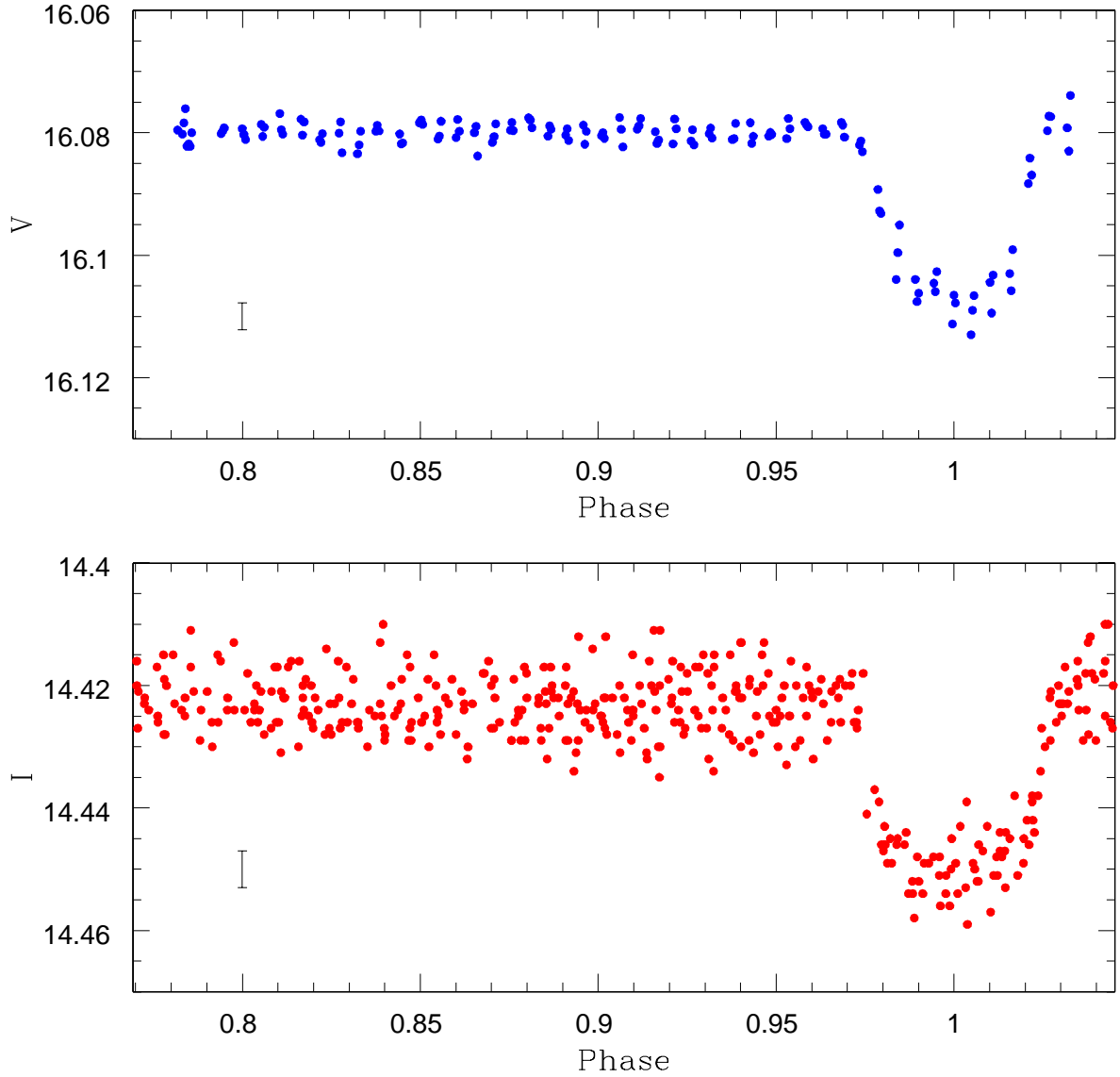}
\caption{Single transit of OGLE-TR-113 observed during April 10th 2005
with VIMOS in the V-band (top) compared with the OGLE phased light
curve transit in the I-band (bottom). Also shown is the size of the
error bars.  The smaller scatter of our photometry is evident.}
\label{completo}
\end{figure}

\begin{figure}
\plotone{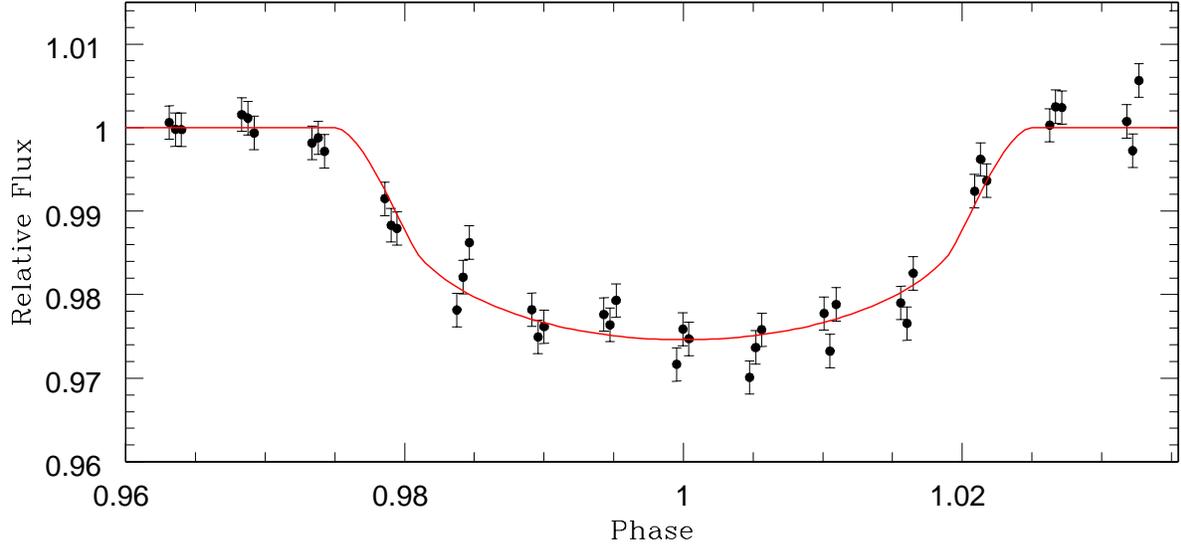}
\caption{
Fit to the single transit of OGLE-TR-113 in the V-band. 
}
\label{fit}
\end{figure}

Figure~\ref{completo} shows the full light curve for the second night
of observations, when the OGLE-TR-113 transit was monitored. For
comparison we also show the phased light curve of the OGLE I-band
photometry (in a similar scale) in the same Figure. The transit is
well sampled in the V-band, and the scatter is smaller.  There are
$\mathrm{N_t} = 30$ points in our single transit, shown in
Figure~\ref{fit}. The minimum is well sampled, allowing us to measure
an accurate amplitude.  In the case of OGLE, the significance of the
transits is --in part-- judged by the number of transits detected. In
the case of the present study, we compute the signal-to-noise of the
single, well sampled transit.  For a given photometric precision of a
single measurement of $\sigma_p$ and a transit depth $\Delta\mathrm{V}$, this
signal-to-noise transit is $S/N=\mathrm{N_t}^{1/2} \Delta\mathrm{V}/\sigma_p$
\citep{gaudi2005}. For OGLE-TR-113 we find the S/N of this transit to
be $S/N\approx76$ using $\Delta\mathrm{V}=0.025$ and $\sigma_p=0.0018$.

The resulting light curve was corrected for a linear trend and phased
adopting a period of P = 1.4324758 days from \citet{konacki2004}. The
curve was fitted using transit curves computed with the algorithms of
\citet{mandel2002}. The curves depend on the ratio between the
planetary and the stellar radii, the ratio between orbit radius and
stellar radius, the inclination angle and the coefficients of the limb
darkening model chosen. All parameters were fitted simultaneously,
with a quadratic model for the limb darkening. We found that
within the errors the limb darkening coefficients remain the same as
those presented by \citet{claret2000} ($u_1 = 0.73$, $u_2 = 0.092$)
for the stellar parameters, which were also employed by
\citet{gillon2006}: $\mathrm{T_{eff}}=4750\mathrm{K}$, metallicity
$\mathrm{[M/H]}=0.0$, surface gravity $\log g = 4.5$ and
microturbulence velocity $\xi_t = 1.0$ km/s. Therefore, the
coefficients were fixed at those values in order to minimize
computation time. The best fit to the transit is shown in
Figure~\ref{fit} as a solid line.

The uncertainties of the fit parameters were estimated from the
$\chi^2$ hypersurface. In the case of non-linear model fitting like
this it is customary to obtain the parameters errors from the
intersection of the $\chi^2$ hypersurface with a constant hyperplane
defined by $\chi^2 = \chi_{bf} + 1$, where $\chi_{bf}$ is the value of
$\chi^2$ associated with the best fit. However, as shown by
\citet{pont2006}, the existence of correlation between the
observations produces a low-frequency noise which must be considered
to obtain a realistic estimation of the uncertainties. To model the
covariance we followed \citet{gillon2006} and obtained an estimate of
the systematic errors in our observations from the residuals of the
light curve. The amplitude of the white ($\sigma_w$) and red
($\sigma_r$) noise can be obtained by solving the following system of
equations:
\begin{eqnarray}
\sigma_1^2 &=&\sigma_w^2 + \sigma_r^2\\
\sigma_8^2 &=&\frac{\sigma_w^2}{8} + \sigma_r^2\; \; ,
\end{eqnarray}
where $\sigma_1$ is the standard deviation taken over single residual
points and $\sigma_8$ is the standard deviation taken over a sliding
average over eight points. We estimated the white noise amplitude,
$\sigma_w = 2.1$ mmag, and the low-frequency red noise amplitude,
$\sigma_r = 0.2$ mmag. 

\begin{figure}
\plotone{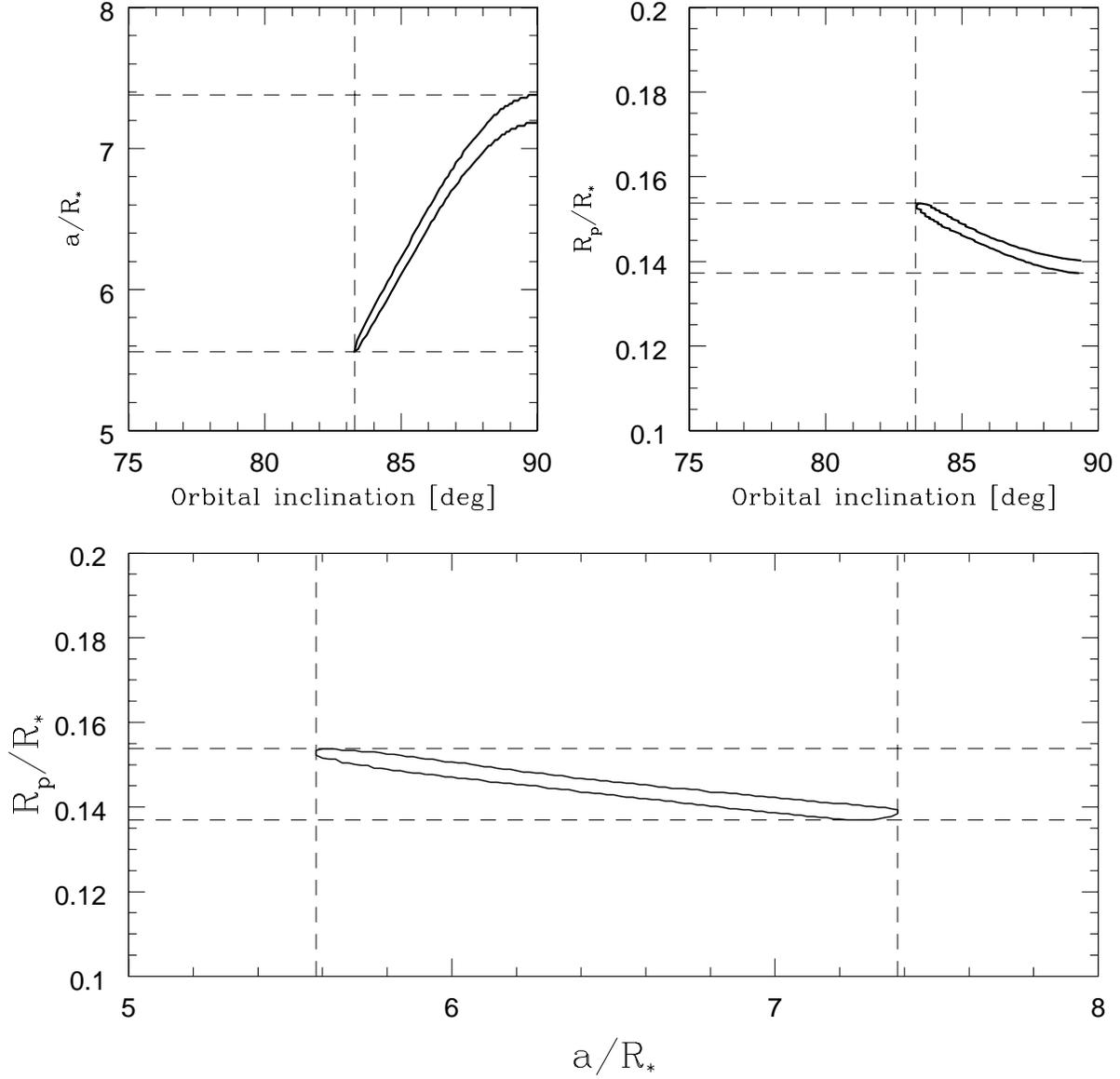}
\caption{ Regions with $\Delta\chi^2\leq1.25$ (see text) for the three
possible projections of the fit parameters. The $\Delta\chi^2\leq1$
regions differ only slightly and are not plotted to avoid confussion.
}
\label{chi}
\end{figure}

Then, we assume that the 1-$\sigma$ uncertainty intervals are determined by
the surface defined by the equation \citep[see][Eq. 7]{gillon2006}:
\begin{equation}
\chi^2 = \chi_{bf} + \Delta\chi^2 = \chi_{bf} + 1 +
\mathrm{N_t}\frac{\sigma_r^2}{\sigma_w^2}\; \; ,
\end{equation}
where $\mathrm{N_t}$ is the number of points in the transit. Using the value
for $\sigma_r$, $\sigma_w$, and $\mathrm{N_t}$ mentioned above, we obtained
$\Delta\chi^2 = 1.25$. The projections of this surface are shown in
Figure~\ref{chi}.
The parameters and uncertainty intervals obtained with this method
are:
\begin{eqnarray}
\mathrm{a/R_*} &=& 6.48 \pm 0.90 \nonumber\\
\mathrm{R_p/R_*} &=& 0.1455 \pm 0.0083  \label{param_fit}\\
i &=& 86.66 \pm 3.34 \nonumber \; \; ,
\end{eqnarray}
where a is the orbit radius, $\mathrm{R_p}$ and $\mathrm{R_*}$ are the
planetary and stellar radius, respectively and $i$ is the orbital
inclination angle.

We also fitted the lightcurve to obtain a refined ephemeris for the
transit times. The parameters obtained above were held fixed and the
time at mean transit was fitted alone. Using the ephemeris presented
by \citet{konacki2004} we also measure a refined period. Our refined
ephemeris for the mean transit times of OGLE-TR113b is:
\begin{displaymath}
HJD(\mathrm{middle~of~transit}) = 2453471.77836(34) +
1.4324752(15)\times E\; \; ,
\end{displaymath}
where the numbers given in parentheses are the uncertainties in the
last digits, about 30 seconds for the central time and 0.2 seconds for
the period. The obtained ephemeris agrees with those from previous
works \citep{konacki2004,gillon2006,snellencovino2006}.

\section{The Radius and other parameters of OGLE-TR-113-b \label{radius}}

There are 20 giant planets with well measured radii to date (November
2006). These are the Solar system planets Jupiter, Saturn, Uranus and
Neptune, plus the extrasolar planets OGLE-TR-10 \citep{konacki2005},
OGLE-TR-56 \citep{konacki2003}, OGLE-TR-111 \citep{pont2004},
OGLE-TR-113 (\citealt{bouchy2004, konacki2004, gillon2006}, and this
work), OGLE-TR-132 \citep{bouchy2004}, HD209458
\citep{charbonneau2000, henry2000}, HD189733 \citep{bouchy2005},
HD149026 \citep{sato2005}, TrES-1 \citep{alonso2004}, TrES-2
\citep{o'donovan2006}, X0-1 \citep{mccullough2006}, HAT-P-1
\citep{bakos2006}, WASP-1 and WASP-2 \citep{colliercameron2006}.
Finally, two more hot Jupiter planets out of 16 bonafide transiting
candidates named SWEEPS-4 and SWEEPS-11 have been discovered by
\citet{sahu2006} in the Galactic bulge.

Our temporal resolution is not enough to obtain an accurate solution
  for all the parameters of the system. In particular, the sampling of
  the ingress and egress is not high enough to allow the breaking of
  the degeneracy between stellar radius and orbital
  inclination. Indeed, assuming a stellar mass of
  $\mathrm{M}_*=(0.78\pm0.02)~\mathrm{M}_\odot$, the error on the
  primary radius obtained from our lightcurve using Kepler's third law
  is around 15\% ($\mathrm{R}_*=(0.76\pm0.12)~\mathrm{R}_\odot$).
  Therefore, to obtain the planetary parameters from the V-band fit
  parameters (see Equations~\ref{param_fit}) we relied on the accurate
  value for stellar radius given by \citet{gillon2006}:
  $\mathrm{R}_*=(0.77\pm0.02)~\mathrm{R}_\odot$. The obtained
  parameters are shown in Table~\ref{resultados}, together with those
  obtained in previous works. Note that due to our poorer temporal
  resolution the uncertainty intervals are larger than those presented
  in previous studies. Nevertheless, there exists an excellent
  agreement with all previous works.

\begin{figure}
\plotone{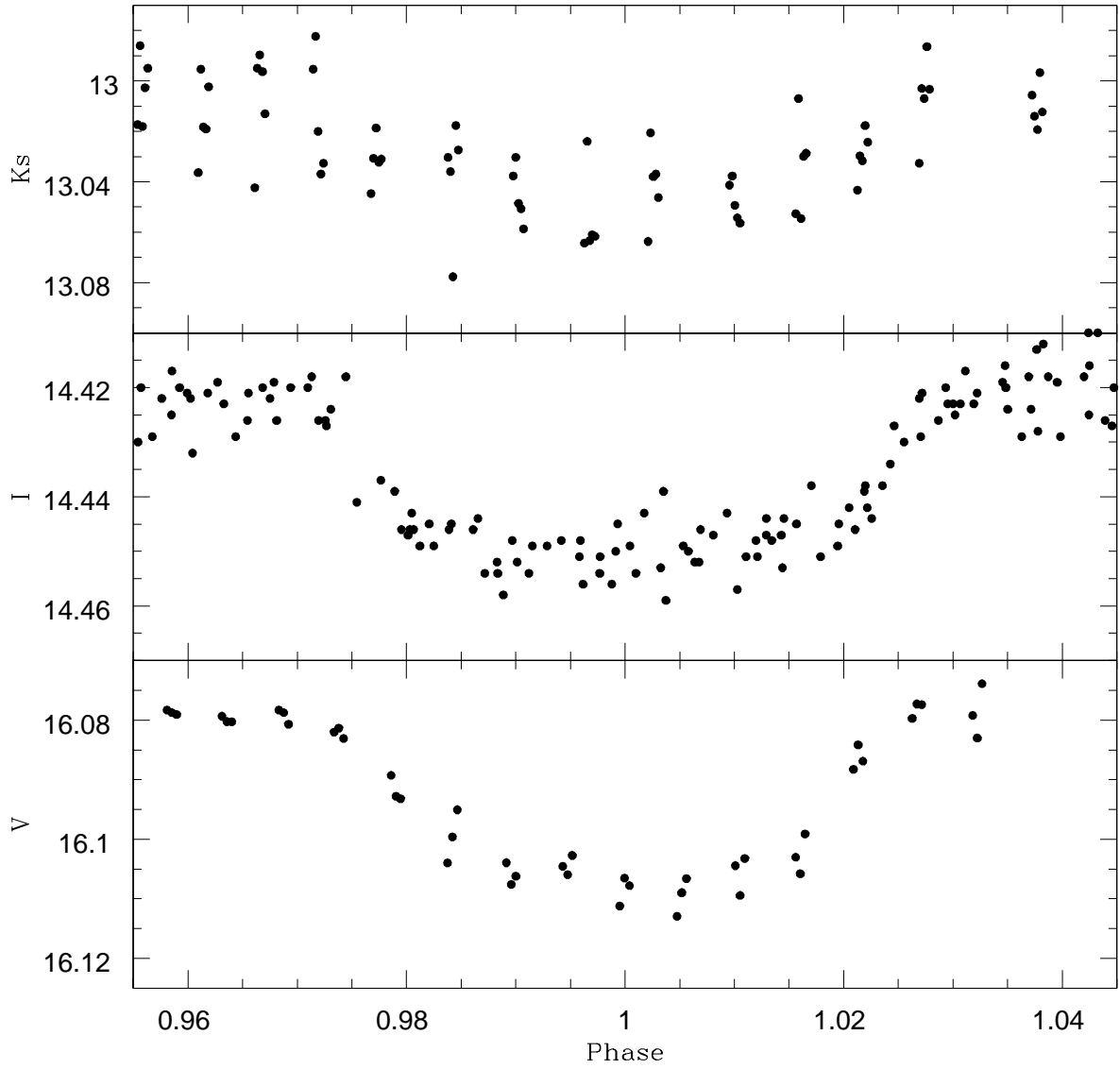}
\caption{ Comparison of an individual transit for OGLE-TR-113 in the
$K_s$-band (top), the phased OGLE I-band light curve for several
transits (middle), and another individual transit in the V-band
(bottom).  The duration and amplitudes of these transits are
consistent with a planet.}
\label{compa_fit}
\end{figure}

In order to measure the radius accurately, we rely on the optical
  photometry. Obviously, the $K_s$-band photometry shows larger
  scatter, and even though we see the whole transit, there is not
  enough baseline covered. Using the IR transit, it is possible to
  derive parameters for the planetary companion to check for
  consistency only.  We fit two lines, one for the out-transit data
  and another for the in-transit one, keeping zero the slope of the
  curve. The difference between the intercepts of each straight line
  equals the depth of the transit. For OGLE-TR-113 we obtained $\Delta
  K_S =0.035\pm0.009$, from which we derive $\mathrm{R_p} =
  1.4\pm0.2$~$\mathrm{R_J}$.  Clearly the error is large, but we find
  a value consistent with the more accurate V-band observations. The
  transits in the I, V and $K_s$ band are shown together for comparison
  in Figure~\ref{compa_fit}.
 
\section{Conclusions}

We studied the OGLE-TR-113 star-planet system using optical and IR
photometry. We have produced an independent confirmation of the dwarf
nature of OGLE-TR-113 by studying the IR and optical CMDs in the
observed fields.

We observed two transits of planet OGLE-TR-113-b. The first was
observed in the V-band with VIMOS at the ESO VLT and the second in the
$K_s$-band with SOFI at the ESO NTT. There are 30 and 40 points per
transit respectively. The quality of the photometry obtained with
VIMOS is superb, with dispersion comparable to photon
noise. Therefore, processing the complete VIMOS dataset should provide
an interesting opportunity to identify new transit candidates around 2
magnitudes fainter than those from the OGLE survey.

We compared our observations with those from previous works
\citep{udalski2003,gillon2006}, and confirmed that the transit
amplitudes in the V, I, and R-bands are similar, consistent with the
planetary nature of the transiting companion.  The $K_s$-band
photometry also shows clearly the transit, although with a large
scatter and insufficient baseline. Within the larger errors, the
parameters are consistent with the more accurate V-band observations.

We checked the limb darkening coefficients using the V-band photometry
  and found the values presented by \citet{claret2000} were
  adequate. However, we were not able to put strong constraints to the
  coefficients in the $K_s$-band, since our light curve is not
  accurate enough.

Finally, the planetary parameters obtained from the V-band photometry
with assumed stellar radius $\mathrm{R}_* = 0.77~\mathrm{R}_\odot$ are
in excellent agreement with those presented in the cited works
\citep{konacki2004,bouchy2004,gillon2006}. We measured the orbit
radius $\mathrm{a}=(0.0232\pm0.0038)$~AU, the orbital period
$\mathrm{P}=(1.4324752\pm0.0000015)$~days, the inclination angle $i =
(86.66\pm3.34)$ degrees and the planetary radius $\mathrm{R_p} =
(1.09\pm0.09)~\mathrm{R_J}$, which for $\mathrm{M_p} =
(1.32\pm0.19)~\mathrm{M_J}$, gives the planet mean density $\rho = (1.26
\pm 0.50)~\mathrm{g~cm^{-3}}$.

\acknowledgements
DM, JMF, GP, MZ, MTR, WG are supported by Fondap
Center for Astrophysics No. 15010003. DM was also supported by a
Fellowship from the John Simon Guggenheim Foundation. AU acknowledges
support from the Polish MNSW DST grant to the Warsaw University
Observatory. We thank the ESO staff at Paranal Observatory.

\clearpage
\begin{table*}
{\scriptsize
\caption{Planetary Parameters}
\begin{tabular} {ccccc}
\hline
\hline
&This work&Guillon et al.(2006)&Bouchy et al.(2004)&Konacki et
al.(2004)\\
\hline
Inclination Angle (deg)&$86.7-90$&$88.8-90$&$85-90$&$88.4\pm2.2$\\
Period (days)&$1.4324752\pm0.0000015$&$1.4324757\pm0.0000013$&$1.43250$ (adopted)&$1.4324758\pm0.0000046$\\
Orbit Radius (AU)&$0.0232\pm0.0038$&$0.0229\pm0.0002$&$0.0228\pm0.0006$&$0.02299\pm0.00058$\\
\\
Planet mass $(\mathrm{M_J})$&$1.32\pm0.19$ (adopted)&$1.32\pm0.19$&$1.35\pm0.22$&$1.08\pm0.28$\\
Planet radius ($\mathrm{R_J}$)&$1.09\pm0.09$&$1.09\pm0.03$&$1.08^{+0.07}_{-0.05}$&$1.09\pm0.10$\\
Planet density (g cm$^{-3}$)&$1.26\pm0.50$&$1.3\pm0.3$&$1.3\pm0.3$&$1.0\pm0.4$\\\hline
\hline
\label{resultados}
\end {tabular}
}
\end {table*}
\clearpage

\end{document}